\documentclass[12pt]{article}
\usepackage[english]{babel}
\usepackage{amsmath}
\usepackage{bbm}
\usepackage{color}
\usepackage[titletoc,toc,title]{appendix}
\Roman{section}
\newcommand{\eq}[1]{\begin{equation}#1\end{equation}}
\newcommand{\naw}[1]{\left(#1\right)}
\newcommand{\ket}[1]{\left|#1\right>}
\newcommand{\bra}[1]{\left<#1\right|}
\newcommand{\av}[1]{\left<#1\right>}
\newcommand{\com}[1]{\left[#1\right]}
\newcommand{\modu}[1]{\left|#1\right|}

\begin{document}

\begin{center}
\textsc{\Large{GATE OPERATORS FOR N-STRATEGIES QUANTUM GAMES}}

\emph{Katarzyna Bolonek-Laso\'n\footnote{kbolonek1@wp.pl}\\ Faculty of Economics and Sociology, Department of Statistical Methods\\ University of Lodz, Poland.}

\end{center}
\begin{abstract}
The quantization of 2-players N-strategies games is considered. The general form of gate operator is determined under the assumption that the classical pure strategies are contained in the set of pure quantum ones. 
\end{abstract}

\section{Introduction}
More than decade ago Eisert, Wilkens and Lewenstein \cite{EisertWL} \cite{EisertW} developed the method which allows, given some classical game, to construct its quantum counterpart. As an example they considered 2-players 2-strategies classical game and proposed their quantum version based on the geometry of $SU(2)$ group. This example provides a paradigm of quantum game. 

To justify the interest in quantum games Eisert, Lewenstein and Wilkens pointed out that there is an intimate connection between the theory of quantum games and the theory of quantum communication. They speculated also that games of survival are being played on molecular level where things are happening according to the laws of quantum mechanics. Moreover, although any quantum game may be modelled by a classical one, it can happen that this is not physically feasible due to limited time and resources; in such a case only quantum mechanics allows for an implementation of the game thanks to the existence of specific correlations which break Bell-like inequalities. Since the appearance of the paper \cite{EisertWL} the theory of quantum games has been a subject of intensive research \cite{BenjaminHay}$\div$\cite{Brunner}.

Two main ingredients of the Eisert et al. method are the choice of the set of admissible strategies and the construction of the gate operator which introduces entanglement in the initial state of the game.

For completeness let us recall the details of ELW game for $N=2$ strategies \cite{EisertWL} \cite{EisertW}.
 The quantization of the game begins by assigning the possible outcomes of the classical strategies C (cooperate) and D ( defect) to the basis vectors $\ket{1}$ and $\ket{2}$ in twodimensional Hilbert space. Depending on their decision each player receives a certain payoff as described in Table 1.

\begin{table}
\caption{The payoffs resulting from different EWL strategies.}
\begin{center}
\begin{tabular}{|c|c|c|c|}\hline
\multicolumn{2}{|c|}{Strategies} & \multicolumn{2}{|c|}{Payoffs}\\
\cline{1-4}
player A & player B  & player A & player B\\
\cline{1-4}
C & C & r & r\\
C & D & s & t\\
D & C & t & s\\
D & D & p & p\\ \hline
\end{tabular}
\end{center}
\end{table}

 The state of the game is described by a vector in the tensor product space spanned by $\ket{1}\otimes\ket{1}$, $\ket{1}\otimes\ket{2}$, $\ket{2}\otimes\ket{1}$ and $\ket{2}\otimes\ket{2}$ which correspond to all possible choices of both players strategies. The initial state of the game is given by
\eq{\Psi_0=\hat{J}\naw{\ket{1}\otimes\ket{1}}}
where $\hat{J}$ is a unitary operator known to both players; $\hat{J}$ represents entaglement and plays an important role in what follows. It is symmetric with respect to the interchange of the players.\\
\indent Strategic moves of the players are associated with unitary operators $\hat{U}_A$, $\hat{U}_B$ operating on their own qubits. The final state of the game is given by 
\eq{\ket{\Psi_f}=\hat{J}^+\naw{\hat{U}_A\otimes\hat{U}_B}\ket{\Psi_0}=\hat{J}^+\naw{\hat{U}_A\otimes\hat{U}_B}\hat{J}\ket{11}}
and the expected payoffs are computed according to
\eq{\begin{split}
& S_A=rP_{11}+pP_{22}+tP_{21}+sP_{12}\\
&  S_B=rP_{11}+pP_{22}+sP_{21}+tP_{12}
\end{split}} 
with $P_{\varepsilon\varepsilon'}\equiv\modu{\av{\varepsilon \varepsilon'|\Psi_f }}^2\equiv\modu{A_{\varepsilon \varepsilon'}}^2$.\\
In order to ensure that the quantum game entails a faithful representation of its classical counterpart (including the mixed strategies of the latter) one selects the two-bit gate operator $\hat{J}$ in the form
\eq{\hat{J}\naw{\gamma}=exp\naw{i\frac{\gamma}{2}\hat{D}\otimes\hat{D}}}
where $\hat{D}=i\sigma_2$ and $\sigma_2$ is the second Pauli matrix.

In the original Eisert et al. proposal the allowed strategies belong to the submanifold of $SU(2)$ group which itself is not a group. This point of view was criticized by Benjamin and Hayden who pointed out that there are no compelling reasons to impose such a restriction. We shall adopt this point of view in what follows.

The gate operator $J$ for the $SU(2)$ case depends on one free parameter $\gamma$. The properties of the game depend, in turn, on the value of this parameter. In the present paper we consider 2-player N-strategies game and construct the multi-parameter general gate operator. We indicate also how the properties of the game depend on the particular choice of $J$. 

\section{The gate operator for N-strategies game}
Our starting point is some classical 2-players N-strategies symmetric game defined by a $N\times N$ payoff matrix $\$^{A,B}$. In order to construct its quantum version one ascribes to any player (Alice and Bob) an N-dimensional complex Hilbert space spanned by the vectors
\begin{equation}
\ket{1}=\left(\begin{array}{c}
1 \\
0 \\
\vdots \\
0 \end{array}\right),\quad \ldots, \quad \ket{N}=\left(\begin{array}{c}
0\\
\vdots\\
0 \\
1
\end{array}\right ).
\end{equation}  

One starts with the vector $\ket{1}\otimes\ket{1}$. The entanglement of the initial state is provided by a reversible gate operator $J$; therefore
\begin{equation}
\ket{\Psi_i}\equiv J\naw{\ket{1}\otimes\ket{1}}
\end{equation}
is the initial state of the game.\\
We assume the set of allowed strategies is the whole $SU(N)$ group. The players perform their moves $U_A$, $U_B$ and then the final measurement is made yielding the final state of the game
\begin{equation}
\ket{\Psi_f}=J^+\naw{U_A\otimes U_B}J\naw{\ket{1}\otimes\ket{1}}.
\end{equation} 
This allows us to compute the players expected payoffs:
\begin{equation}
\$^{A,B}=\sum_{\sigma,\sigma'=1}^{N}p_{\sigma,\sigma'}^{A,B}\modu{\av{\sigma,\sigma'|\Psi_f}}^2
\end{equation}
where $\ket{\sigma,\sigma'}\equiv\ket{\sigma}\otimes\ket{\sigma'}$ and $p_{\sigma,\sigma'}^{A,B}$ are classical payoffs of Alice and Bob, respectively.
In order to construct a gate operator $J$ we assume that all classical pure strategies are contained in the set of pure quantum ones. A sufficient condition for this to be the case is the existence of $N$ matrices $U_\sigma\in SU(N)$, $\sigma=1,...,N$ such that: (i) $U_\sigma\ket{1}=e^{i\varphi_\sigma}\ket{\sigma}$, $\sigma=1,...,N$; (ii) $\com{J,U_\sigma\otimes U_{\sigma'}}=0$, $\sigma, \sigma'=1,...,N$. To leave as much freedom as possible for the choice of $J$ we assume further that
\begin{equation}
\com{U_\sigma,U_{\sigma'}}=0,\quad \sigma,\sigma'=1,...,N.\label{a}
\end{equation} 
In order to construct the matrices $U_\sigma$ it is sufficient to consider the representation of the subgroup of cyclic permutations of $12...N$. To this end consider the matrix
\begin{equation}
U=\left(\begin{array}{ccccc}
0 & 0 & \cdots\cdots & 0&1 \\
1 & 0 & \cdots\cdots & 0 &0\\
0 & 1 & \cdots\cdots & 0 &0\\
\cdots & \cdots &\cdots\cdots& \cdots&\cdots\\
0 & 0 & \cdots\cdots & 1 & 0 
\end{array}\right).
\end{equation}
Then
\begin{equation}
\begin{split}
& U\ket{\sigma}=\ket{\sigma+1},\quad \sigma=1,...,N-1 \\
& U\ket{N}=\ket{1}\\
& U^N=\mathbbm{1} \\
& \det U=\naw{-1}^{N-1}. 
\end{split}
\end{equation}
Let us define
\begin{equation}
U_\sigma=e^{\frac{i\pi\naw{N-1}\naw{\sigma-1}}{N}}U^{\sigma-1},\qquad \sigma=1,...,N.
\end{equation}
Then (\ref{a}) and (i), with $\varphi_{\sigma}=\frac{\pi\naw{N-1}\naw{\sigma-1}}{N}$, are obeyed.

The eigenvalues of $U$ are $1,\varepsilon, \varepsilon^2,...,\varepsilon^{N-1}$ with $\varepsilon=\exp\naw{\frac{2i\pi}{N}}$ being the first primitive $N$-th root of unity. It is not difficult to find the corresponding eigenvectors and the matrix $V$ diagonalizing $U$; the latter reads
\begin{equation}
V_{ik}=\frac{1}{\sqrt{N}}\overline{\varepsilon}^{\naw{i-1}\naw{k-1}},\qquad i,k=1,...,N.
\end{equation}
The necessary and sufficient condition for (ii) to hold is
\begin{equation}
\com{J,\mathbbm{1}\otimes U}=0=\com{U\otimes\mathbbm{1},J}.
\end{equation}
Let us define
\begin{equation}
\widetilde{J}=\naw{V^+\otimes V^+}J\naw{V\otimes V}.
\end{equation}
Due to the equality $V^+UV=diag\naw{1,\varepsilon,...,\varepsilon^{N-1}}$ $\widetilde{J}$ must be diagonal. Let $\Lambda_i$, $i=1,...,N-1$ be any basis in Cartan subalgebra of $SU(N)$.\\
Then $\widetilde{J}$ can be written as
\begin{equation}
\widetilde{J}=\exp\naw{i\sum_{k=1}^{N-1}\lambda_k\naw{\Lambda_k\otimes\Lambda_k}+i\sum_{k\neq l=1}^{N-1}\mu_{kl}\naw{\Lambda_k\otimes\Lambda_l+\Lambda_l\otimes\Lambda_k}}\label{a1}
\end{equation}
with $\mu_{kl}=\mu_{lk}$ and
\begin{equation}
J=\naw{V\otimes V}\widetilde{J}\naw{V^+\otimes V^+}.\label{a2}
\end{equation}
In defining $\widetilde{J}$ we omitted in the exponent the term $\mathbbm{1}\otimes\mathbbm{1}$ (it gives an irrelevant phase) as well as the terms $\mathbbm{1}\otimes \Lambda_k+\Lambda_k\otimes\mathbbm{1}$ (which amount to relabelling of the set of strategies).

Eqs. (\ref{a1}) and (\ref{a2}) provide the expression for gate operator. It depends on $N-1+{N-1 \choose 2}={N \choose 2}$ free parameters.

The above construction can be further generalized by replacing the matrix $U$ by a more general one
\begin{equation}
U=\left(\begin{array}{ccccc}
0 & 0 & \cdots\cdots & 0& e^{i\varphi_N} \\
e^{i\varphi_1} & 0 & \cdots\cdots & 0 &0\\
0 & e^{i\varphi_2} & \cdots\cdots & 0 &0\\
\cdots & \cdots &\cdots\cdots& \cdots&\cdots\\
0 & 0 & \cdots\cdots & e^{i\varphi_{N-1}} & 0 
\end{array}\right)
\end{equation}
and repeating the above reasoning with appropriate modifications. For example, the original ELW game $(N=2)$ is recovered with $\varphi_1=\pi$, $\varphi_2=0$.

\section{Conclusions}
We have constructed the gate operators for quantum versions of classical 2-players N-strategies symmetric game. The basic assumption was that the classical pure strategies are contained in the set of pure quantum ones. In the original ELW game $(N=2)$ the pure quantum strategies contain also all mixed classical one. For general $N$ it is no longer the case \cite{Bolonek1}. However, mixed quantum strategies contain also mixed classical ones.

Once the gate operator is determined one can study the entanglement of initial state and its influence on the properties of our game. The degree of entanglement is determined by the structure of the reduced density matrix
\begin{equation}
\rho_{red}=\text{Tr}_B\rho_i\quad \naw{or\, \text{Tr}_A\rho_i},\quad \rho_i=\ket{\Psi_i}\bra{\Psi_i}.
\end{equation} 
In particular, $\rho_{red}=\frac{1}{N}\mathbbm{1}$ corresponds to maximal entanglement. It is not difficult to show \cite{Bolonek2} that in such case to any strategy there exists an appropriate counterstrategy \cite{Bolonek2}. As a result no pure nontrivial Nash equilibrium exists. The condition of the maximal entanglement of initial state imposes restrictions on the values of free parameters. In this way one identifies the set of gate operators which lead to the quantum games with no nontrivial pure Nash equilibria.

The case of maximal entanglement corresponds to the maximal stability subgroup of initial state vector. It is isomorphic to $SU(N)$ leaving the set of effective strategies for $\underline{\text{both}}$ players restricted to $SU(N)$ manifold ($SU(N)\times SU(N)/SU(N)\sim SU(N)$ topologically). One can go further and study the cases of partial entanglement by considering the degeneracy pattern of reduced density matrix. The latter determines the stability subgroup and, as a result, the dimension of the manifold of effective strategies of both players. The $N=3$ example is considered in some detail in Ref. \cite{Bolonek2}.

\subsection*{Acknowledgement}
I would like to thank Professor Piotr Kosi\'nski (Department of Computer Science, Faculty of Physics and Applied Informatics, University of L\'od\'z, Poland) for helpful discussion and useful remarks. 
 This research is supported by the NCN Grant no. DEC-2012/05/D/ST2/00754.

\end{document}